\documentclass[12pt]{article}
\usepackage{fullpage,amsmath}
\usepackage[pdftex,dvipsnames,usenames]{color}         %for colors
\usepackage{xcolor}
\usepackage{natbib,bm}
\usepackage{float}
\usepackage{lscape}
\usepackage{bbm}
\usepackage{graphicx,psfrag,epsf}
\usepackage{enumerate}
\usepackage{enumitem}
\usepackage{natbib}
\usepackage{bbm} 
\usepackage{amsmath,amsthm,amssymb,bm}
\usepackage{multirow}

\usepackage{abstract}
\usepackage{amsthm} %to include command ``\newtheorem*`` 
\newcommand{\comment}[1]{}

\def\boldfacefake#1{\kern-4pt
   \hbox{ \mathsurround=0pt
   \hbox to 0.4pt{$#1$\hss}\hbox to 0.4pt{$#1$\hss}\hbox {$#1$}}}

\newcommand{\ba}{\begin{eqnarray*}}
\newcommand{\ea}{\end{eqnarray*}}

\newtheorem{theorem0}{Theorem}
\newtheorem{lemma0}{Lemma}
\newtheorem{remark0}{Remark}
\newtheorem{fact0}{Fact}
\newtheorem{example0}{Example}
\newtheorem{definition0}{Definition}
\newtheorem{corollary0}{Corollary}
\newtheorem{proposition0}{Proposition}
\newtheorem{algorithmY}{Algorithm}
\newtheorem{condition0}{Condition}
\newtheorem{assumption0}{Assumption}
\newtheorem{simulation0}{Simulation}

\newcommand{\reals}{\mbox{\rm I\kern-.20em R}}
\newcommand{\sreals}{\mbox{\small \rm I\kern-.20em R}}

\newcommand{\bqg}{\begin{quote} \color{Green}\em}%X
	\newcommand{\toale}{\end{quote} \color{black}\rm}
\newcommand{\eqg}{\end{quote} \color{black}\rm}%X
\newcommand{\bd}{\begin{description}}
\newcommand{\ed}{\end{description}}
\newcommand{\bi}{\begin{itemize}}
\newcommand{\ei}{\end{itemize}}
\newcommand{\be}{\begin{enumerate}}
\newcommand{\ee}{\end{enumerate}}

\usepackage[normalem]{ulem}
 
\begin{document}
 
% --------------------------------------------------------------
%                         Start here
% --------------------------------------------------------------
%need title page
	\begin{center}
		{\large \bf A Simulation Study of the Performance of Statistical Models for Count Outcomes with Excessive Zeros}
		
		\vspace{.7 cm}
		
		Zhengyang Zhou, Ph.D. \\
		Department of Biostatistics and Epidemiology \\
		University of North Texas Health Science Center, Fort Worth, TX \\
		
		\vspace{.5 cm}
		Dateng Li, Ph.D. \\
		121 Westmoreland Ave., White Plains, NY \\
		
				\vspace{.5 cm}
		David Huh, Ph.D.  \\
		School of Social Work \\
		University of Washington,	Seattle, WA\\
		
		\vspace{.5 cm}
		Minge Xie, Ph.D. \\
		Department of Statistics\\
		Rutgers University,
		Piscataway, NJ \\

	    \vspace{.5 cm}
		Eun-Young Mun, Ph.D.  \\
		Department of Health Behavior and Health Systems \\
		University of North Texas Health Science Center, Fort Worth, TX \\
		
	\end{center}
	
	\vspace{1cm}
	
	\noindent
	Correspondence should be sent to:\\
		Zhengyang Zhou, Ph.D. \\
		Email: zhengyang.zhou@unthsc.edu 
 \\
	\title{}
	\author{}
	\date{}
	\maketitle
	\begin{abstract}
		\begin{center}
		{\large \bf Abstract
		}
		\end{center}

\noindent \textbf{Background}:  Outcome measures that are count variables with excessive zeros are common in health behaviors research. Examples include the number of standard drinks consumed or alcohol-related problems experienced over time. There is a lack of empirical data about the relative performance of prevailing statistical models for assessing the efficacy of interventions when outcomes are zero-inflated, particularly compared with recently developed marginalized count regression approaches for such data.\\
%When these count outcomes are evaluated in various statistical models to assess the efficacy of interventions, their relative performance is unknown, especially compared to novel models. \\
\textbf{Methods}: The current simulation study examined five commonly used approaches for analyzing count outcomes, including two linear models (with outcomes on raw and log-transformed scales, respectively) and three prevailing count distribution-based models (i.e., Poisson, negative binomial, and zero-inflated Poisson (ZIP) models). We also considered the marginalized zero-inflated Poisson (MZIP) model, a novel alternative that estimates the overall effects on the population mean while adjusting for zero-inflation. Motivated by alcohol misuse prevention trials, extensive simulations were conducted to evaluate and compare the statistical power and Type I error rate of the statistical models and approaches across data conditions that varied in sample size ($N =$ 100 to 500), zero rate (0.2 to 0.8), and intervention effect sizes.\\ %Different intervention effects on reducing alcohol consumption, proportions of zero standard drinks, and trial sample size (i.e., 100, 200, 300, 500) were considered in the simulation study.\\
\textbf{Results}: Under zero-inflation, the Poisson model failed to control the Type I error rate, resulting in  higher than expected false positive results. When the intervention effects on the zero (vs. non-zero) and count parts were in the same direction, the MZIP model had the highest statistical power, followed by the linear model with outcomes on the raw scale, negative binomial model, and ZIP model. The performance of the linear model with a log-transformed outcome variable was unsatisfactory.\\ %When only one of the effects on the zero (vs. non-zero) part and the count part existed, the ZIP model had the highest statistical power.     \\
\textbf{Conclusions}: The MZIP model demonstrated better statistical properties in detecting true intervention effects and controlling false positive results for zero-inflated count outcomes. This MZIP model may serve as an appealing analytical approach to evaluating overall intervention effects in studies with count outcomes marked by excessive zeros. 
	\end{abstract}
	

\date{}
\maketitle

\section{Introduction}
Count outcomes are frequently encountered in health behaviors research. Examples of such data include number of standard drinks consumed \citep{huh2019tutorial}, number of cigarettes smoked \citep{sheu2004effect}, and number of sexual risk behaviors experienced \citep{hutchinson2003role}. Zero-inflation occurs when there is an excessive proportion of outcome values stacked at zero, which is a common phenomenon, especially with behavioral risk outcomes. For example, in alcohol prevention and intervention trials aimed at reducing alcohol consumption among participants, the proportion of participants reporting zero alcohol drink can be as high as 66\%, suggesting that the outcome variable was zero inflated \citep{huh2015brief}. One reason for the disproportionate proportions of zeros and non-zero values is that the study population may consist of two clinically distinct groups, where one group is represented by participants at-risk for alcohol consumption and the other by participants not-at-risk (e.g., participants who are abstainers that will not consume any drink, resulting in zero-inflation). In the following context, we refer to the above two groups as  the \textit{at-risk} and \textit{not-at-risk} subpopulations, respectively.

Many types of statistical approaches have been utilized to model count data in the literature. To appropriately account for the count nature of data, researchers have used generalized linear models based on count distributions, such as the Poisson and negative binomial (NB). The Poisson regression model assumes that the mean of the outcome is equal to the variance, while the NB regression model allows the variance to be greater than the mean by incorporating an additional dispersion parameter. Both the Poisson and NB models assume that the study sample comes from one homogeneous population and relate covariates to the mean outcome of the entire sample. With these models, there is no flexibility to account for excessive zeros when the count outcome of interest is zero inflated. 

The zero-inflated Poisson (ZIP) model is an extension of the regular Poisson model that is more appropriate for count data with excessive zeros by using a mixture
distribution of the Poisson distribution and a point mass at zero (i.e., the structural zeros). In the context of alcohol intervention trials, the Poisson part can be considered as evaluating the \textit{at-risk} subpopulation who may or may not drink at a given assessment, and the structural zero part as evaluating the \textit{not-at-risk} subpopulation who ``predictably'' do not drink (e.g., abstainers for religious or other personal reasons). More discussion for the two-part nature of the ZIP model can be found in \cite{zhou2021bias}. Unlike the Poisson and NB models that evaluate the effects of each covariate on the overall mean of the outcome, the ZIP model separately evaluates the effects on the two parameters of the mixture distribution--the Poisson mean and the probability of a structural zero. As a consequence, the estimates from a ZIP model can be cumbersome to interpret, as they describe two different parameters for two subpopulations. 

To directly infer the effects on the overall mean and maintain the ability to account for zero inflation, the marginalized ZIP (MZIP) model has been proposed based on the framework of the ZIP model \citep{long2014marginalized,preisser2017matching,famoye2018marginalized}. Instead of separately evaluating the effects on the two parameters, this approach makes direct
inference on the overall effect of the entire sample by linking the marginal (or overall) mean of the outcome to the covariates. Compared to the ZIP model, which conceptually separates the population into two subpopulations, the MZIP model treats the entire sample as a whole, which makes it feasible to answer the following, simpler but often critical question of \textit{whether the intervention is efficacious for the entire study population}. That question is commonly of principal interest when a clinical trial is designed \citep{mun2022brief} and can be accounted for in a calculation of sample size based on an MZIP model \citep{zhou2022sample}. %\textcolor{red}{Following the introduction of the MZIP model, other methods directly relating covariates to the overall mean have been proposed based on the idea of mean marginalization (e.g., \cite{todem2016marginal,preisser2016marginalized}).}

Despite the increasing availability of new statistical methods and software for analyzing count data with zero inflation, a nonignorable number of studies still do not utilize appropriate statistical methods for such data. For example, in a meta-analysis of 17 studies using individual participant data from each, over half (nine studies) had excessive proportions of zero outcomes (i.e., number of drinks). However, of the nine, eight did not account for this zero inflation \citep{huh2015brief}. A review by \cite{tan2022statistical} summarized the statistical models used to evaluate the effectiveness of brief alcohol interventions in reducing alcohol consumption. The investigators reviewed 119 alcohol-related count outcomes from 64 papers and observed that more than half of the outcomes (61.3\%) were analyzed using statistical models that assume normally distributed residuals. Less than a third (31.1\%) were analyzed using count distribution models. These observations suggest a gap between the methodological advances and their applications in applied research. 

%The reasons for this gap are two-fold. First, there is a lack of guidance in the model selection among the available count- and normal-based models under a large range of possible data conditions, which can assist researchers with selecting a more appropriate statistical approach to analyzing their data. Second, \textcolor{red}{although novel statistical models for count data have been proposed over time, some of them have not been disseminated in scientific fields outside of statistics and biostatistics}, hindering their implementation for research applications. For example, the MZIP model was proposed in 2014 \citep{long2014marginalized} and was applied for the first time in alcohol intervention research in 2022 \citep{mun2022brief} --- eight years following its introduction. 

%\sout{The reasons for this gap are two-fold. First, there is a lack of guidance in the model selection among the available count- and normal-based models under a large range of possible data conditions, which can assist researchers with selecting a more appropriate statistical approach to analyzing their data. Second, although novel count statistical models have been proposed in statistics, they have remained invisible or inaccessible in scientific fields outside of statistics and biostatistics, hindering their implementation for research applications. For example, the MZIP model was proposed in 2014 \citep{long2014marginalized} and was applied for the first time in alcohol intervention research in 2022 \citep{mun2022brief} — eight years following its introduction.}

In this article, we aim to bridge the implementation gap by providing evidence-based guidance in selecting appropriate statistical models for count data with or without excessive zeroes through extensive simulation studies. We evaluate Type I error (i.e., ability to control false positives) and statistical power (i.e., ability to detect true effects) of candidate methods for count data under a large range of conditions. More specifically, we consider \textcolor{black}{three broad sets of statistical models. The first set consists of conventional count-distribution based models for data with or without zero-inflation, including the Poisson, NB, and ZIP models. The second is a marginalized model for zero-inflated data, and we consider the MZIP model. The third set consists of linear models, with or without logarithm transformation, which have been commonly used in the literature (e.g., see \cite{tan2022statistical}).} Notably, this is the first study to systematically evaluate the statistical properties of the MZIP model and compare it to other alternative approaches.

This article is organized as follows. In Section 2, we describe the formulation of the count distribution-based models considered in the simulation study. In Section 3, we describe a simulation study to evaluate and compare the relative performances of candidate methods under various data conditions. In Section 4, we summarize the empirical results in terms of Type I error and statistical power obtained from the simulation study. In Section 5, we discuss the overall findings and conclusions.

\section{Methods}
For a clinical trial with two arms, let us assume that the outcome of interest is a count variable that may or may not have excessive zero values. Suppose that the study sample size is $n$ and for the $i$-th participant, $i=1,2,...,n$, the count outcome is $y_i$. Consider $p-1$ covariates in the statistical model of the trial outcome, one of which is the intervention assignment indicator $\mathbbm{1}_{\{A_i=T\}}$, where $A_i$ denotes $i$-th participant's assignment to either the intervention ($T$) or control ($C$) arm. Denote the remaining $p-2$ covariates as ${\bf x}_{i,p-2}=(x_{i2},x_{i3},...,x_{i,p-1})^t$. In the following, we describe potential statistical models that may be considered to evaluate the intervention effect on the count outcome, including the Poisson, NB, ZIP, MZIP, and linear regression models with raw scale  scores or log-transformed scores.

\subsection{Poisson and NB regression models}

Among the count distribution-based regression models, the Poisson model has the most straightforward formulation by modeling the logarithm of the mean outcome through a list of predictors. The outcome values are assumed to follow the Poisson distribution, which restricts the mean value to be equal to its variance. When there is ``overdispersion" in the data, where the variance of the distribution is larger than the mean, the Poisson model can underestimate variance and yield invalid inferences. Based on the two-arm trial design described at the beginning of the Methods Section, the Poisson model can be expressed as
\begin{equation}
\begin{split}
\label{eq:poisson}
&y_i\sim \text{Poisson}(v_i), \\
&\log(v_i)={\bf x}_i^t{\bm \beta^{Poi}}=\beta_0^{Poi}+\beta_1^{Poi}\mathbbm{1}_{\{A_i=T\}}+{\bf x}_{i,p-2}^t{\bm \eta^{Poi}},
\end{split}
\end{equation}
where $v_i=\mathbb{E}[y_i]$ is the overall mean of the outcome under a Poisson distribution, ${\bf x}_i=(1, \mathbbm{1}_{\{A_i=T\}}, {\bf x}_{i,p-2}^t)^t$, and ${\bm \beta^{Poi}}=(\beta_0^{Poi},\beta_1^{Poi},\bm \eta^{Poi(t)})^t=(\beta_0^{Poi},\beta_1^{Poi},\beta_2^{Poi},...,\beta_{p-1}^{Poi})^t$ are the vectors of regressors and regression coefficients, respectively. 

The NB model is an alternative count regression model. Compared to the Poisson regression model, it incorporates an additional ``dispersion'' parameter, which allows the variance to be greater than the mean. Therefore, the NB regression model is flexible in accommodating overdispersion. Similarly, the NB regression model can be expressed as
\begin{equation}
\begin{split}
\label{eq:nb}
&y_i\sim \text{NB}(v_i,k), \\
&\log(v_i)={\bf x}_i^t{\bm \beta^{NB}}=\beta_0^{NB}+\beta_1^{NB}\mathbbm{1}_{\{A_i=T\}}+{\bf x}_{i,p-2}^t{\bm \eta^{NB}},
\end{split}
\end{equation}
where $v_i=\mathbb{E}[y_i]$ is the overall mean of the outcome and $k > 0$ is the dispersion parameter, which satisfies $\mathrm{Var}[y_i] = v_i + \frac{v_i^2}{k}$. \textcolor{black}{Of note, the above parameterization for dispersion follows the Type 2 NB distribution (or NB2) with a quadratic mean-variance relationship \citep{hilbe2011negative}. For a Type 1 NB distribution parameterization, the variance is a linear function of the mean, which is less commonly used in practice.}

\subsection{ZIP regression model}
When evaluating count outcomes, zero inflation is present when the observed proportion of zeros is much greater than the theoretically expected proportion under a conventional count distribution, such as Poisson or NB. In the presence of zero inflation, the Poisson and NB models may not perform well because their formulations do not account for excessive zeros. As a result, the two statistical models (i.e., Poisson and NB models) could produce biased effect size estimates and inaccurate statistical significance inferences \citep{perumean2013zero}.

The ZIP model was proposed to account for zero inflation by explicitly modeling excessive zeros \citep{mullahy1986specification, lambert1992zero}. This model assumes that a count outcome follows a mixture distribution consisting of the Poisson distribution and a point mass at zero (i.e., the structural zeros). For example, in the context of alcohol intervention trials whose primary goal is to reduce the number of drinks consumed, the Poisson part can be considered as evaluating an \textit{at-risk} subpopulation who may or may not drink at a given assessment. In contrast, the structural zero part can be considered as evaluating the \textit{not-at-risk} subpopulation who ``predictably'' do not drink (e.g., abstainers for religious or other personal reasons). More discussion on the two-part nature of the ZIP model can be found in \cite{zhou2021bias}. Consequently, the intervention effects are estimated in two separate parts -- the rate ratio (RR) of the mean in the Poisson part (e.g., number of drinks, including random zeros from those who happened not to drink) and the odds ratio (OR) of being a structural zero (e.g., abstainers vs. non-abstainers) in the structural zero part. The ZIP model can be formally expressed as
\begin{equation}
\begin{split}
\label{eq:ZIP}
  &y_i\sim \begin{cases}
               0 &\text{with probability } \pi_i\\
               \text{Poisson($\mu_i$)} &\text{with probability }1-\pi_i
            \end{cases},\\
  &\log(\frac{\pi_i}{1-\pi_i})={\bf x}_i^t{\bm \gamma^{ZIP}}=\gamma_0^{ZIP}+\gamma_1^{ZIP}\mathbbm{1}_{\{A_i=T\}}+{\bf x}_{i,p-2}^t{\bm \zeta^{ZIP}} \text{, and}\\
  &\log(\mu_i)={\bf x}_i^t{\bm \beta^{ZIP}}=\beta_0^{ZIP}+\beta_1^{ZIP}\mathbbm{1}_{\{A_i=T\}}+{\bf x}_{i,p-2}^t{\bm \eta^{ZIP}},
\end{split}
\end{equation}
\sloppy where $\mu_i=\mathbb{E}[y_i|y_i \text{ from the Poisson part}]$ is the mean parameter of the Poisson part, $\pi_i=Pr(y_i \text{ is a structural zero})$ is the structural zero rate, ${\bm \beta^{ZIP}}=(\beta_0^{ZIP},\beta_1^{ZIP},\bm \eta^{ZIP(t)})^t=(\beta_0^{ZIP},\beta_1^{ZIP},\beta_2^{ZIP},...,\beta_{p-1}^{ZIP})^t$ are the regression coefficients for the Poisson part, and ${\bm \gamma^{ZIP}}=(\gamma_0^{ZIP},\gamma_1^{ZIP},\bm \zeta^{ZIP(t)})^t=(\gamma_0^{ZIP},\gamma_1^{ZIP},\gamma_2^{ZIP},...,\gamma_{p-1}^{ZIP})^t$ are the regression coefficients for the structural zero part. When applying the ZIP model to data, the intervention effects are evaluated separately for the two parts, corresponding to two distinct subpopulations. However, in many clinical trials, whether there is an ``overall" intervention effect for the entire population is an important clinical question. Unfortunately, the overall intervention effect is not straightforward to evaluate in a ZIP model. %${\bm \beta}=(\beta_0,\beta_1,\bm \eta^t)^t=(\beta_0,\beta_1,\beta_2,...,\beta_{p-1})^t$ are the regression coefficients for the Poisson part, and ${\bm \gamma}=(\gamma_0,\gamma_1,\bm \zeta^t)^t=(\gamma_0,\gamma_1,\gamma_2,...,\gamma_{p-1})^t$ are the regression coefficients for the structural zero part.
Of note, under the ZIP model, the ``overall mean," which is denoted by $\mathbb{E}[y_i]\triangleq  v_i$, can be expressed as  
\begin{equation}
\label{eq:ZIP2}
v_i = (1-\pi_i)\mu_i = \frac{e^{{\bf x}_i^t{\bm \beta^{ZIP}}}}{1+e^{{\bf x}_i^t{\bm \gamma^{ZIP}}}}.
\end{equation}
Equation (\ref{eq:ZIP2}) implies that the ``overall mean" depends on all covariates and consequently all parameters from the two parts of the model.  
More importantly, the overall effect of the intervention, which is usually defined as the incidence rate ratio (IRR) between the intervention (T) and control (C) groups holding other covariates constant, is expressed as
\begin{equation}
\label{TE:ZIP}
\frac{\mathbb{E}[y_i|A_i=T,{\bf x}_{p-2}]}{\mathbb{E}[y_j|A_j=C,{\bf x}_{p-2}]} = \exp(\beta_1^{ZIP})\frac{1+\exp({\gamma_0^{ZIP}+{\bf x}_{p-2}^t{\bm \zeta^{ZIP}}})}{1+\exp{(\gamma_0^{ZIP} + \gamma_1^{ZIP}+{\bf x}_{p-2}^t{\bm \zeta^{ZIP}}})},
\end{equation}
where $i$ and $j$ represent two hypothetical participants in treatment and control groups, respectively, and with the same sets of covariates (i.e., ${\bf x}_{p-2}$).
Equation (\ref{TE:ZIP}) implies that under the ZIP model, unless the treatment indicator is the only covariate included in the model, the intervention effect varies across individuals and depends on 
all other covariates. To obtain a population-level overall treatment effect from the ZIP model, it is necessary to integrate out all covariates, which can be computationally tedious and error prone.

\subsection{MZIP model}
The MZIP model is an extension of the ZIP model \citep{long2014marginalized,preisser2017matching,famoye2018marginalized}. The MZIP model accounts for zero inflation and directly models the overall mean of the outcome. Recall that we denote $\mathbb{E}[y_i]\triangleq v_i$ as the overall mean and $\mu_i$ as the mean of the Poisson variable. The MZIP model is then expressed as
\begin{equation}
\begin{split}
\label{eq:MZIP}
  &y_i\sim \begin{cases}
               0 &\text{with probability } \pi_i\\
               \text{Poisson($\mu_i$)} &\text{with probability }1-\pi_i
            \end{cases},\\
  &\log(\frac{\pi_i}{1-\pi_i})={\bf x}_i^t{\bm \gamma^{MZIP}}=\gamma_0^{MZIP}+\gamma_1^{MZIP}\mathbbm{1}_{\{A_i=T\}}+{\bf x}_{i,p-2}^t{\bm \zeta^{MZIP}} \text{, and}  \\
  &\log(v_i)={\bf x}_i^t{\bm \beta^{MZIP}}=\beta_0^{MZIP}+\beta_1^{MZIP}\mathbbm{1}_{\{A_i=T\}}+{\bf x}_{i,p-2}^t{\bm \eta^{MZIP}}.
\end{split}
\end{equation}
Note that the MZIP model is different from the ZIP model in that the MZIP \textit{directly} models the overall mean (i.e., $v_i$) through covariates, instead of the mean in the Poisson part (i.e., $\mu_i$) as in the ZIP model (see Equation (4)). No additional regression equation is needed for the Poisson mean, $\mu_i$, as it is determined through the equation $\mu_i = \frac{v_i}{1-\pi_i}$.

In Equation (\ref{eq:MZIP}), the intervention effect on the ``overall mean" outcome for the entire population is quantified by $\beta_1^{MZIP}$, which can be interpreted as the log incidence density ratio difference between the intervention and control groups. Therefore, $\beta_1^{MZIP}$ enjoys the same straightforward interpretation as $\beta_1^{Poi}$ or $\beta_1^{NB}$. With a single intervention effect estimate on the entire population, the MZIP model is advantageous over the ZIP model when answering the question of whether, and to what extent, an intervention in question is efficacious for the entire population. In addition to the intervention effect on the overall mean estimate, the MZIP model provides the parameter estimates in the structural zero part using the same formula as in the ZIP model.

The likelihood function of an MZIP model is as follows:
\begin{equation}
\begin{split}
\label{eq:likelihood}
L(\bm \gamma^{MZIP},\bm \beta^{MZIP}|\bm y) & = \prod_{y_i}(1+e^{{\bf x}_i^t{\bm \gamma^{MZIP}}})^{-1}\prod_{y_i=0}[e^{{\bf x}_i^t{\bm \gamma^{MZIP}}} + e^{-(1+\exp{({\bf x}_i^t{\bm \gamma^{MZIP}}))\exp{({\bf x}_i^t{\bm \beta^{MZIP}})}}}]\\
&  \prod_{y_i > 0}\frac{[e^{-(1+\exp{({\bf x}_i^t{\bm \gamma^{MZIP}}))\exp{({\bf x}_i^t{\bm \beta^{MZIP}})}}}(1+e^{{\bf x}_i^t{\bm \gamma^{MZIP}}})^{y_i}e^{y_i{\bf x}_i^t{\bm \beta^{MZIP}}}]}{y_i!}
\end{split}
\end{equation}
To fit an MZIP model, one can estimate the parameters by maximizing the likelihood function shown in Equation (\ref{eq:likelihood}) using non-linear optimization algorithms. To disseminate the application of the MZIP model, our group has developed an R package, ``mcount'' \citep{mcount} that can fit an MZIP model (see \cite{mun2022brief} for a real data application utilizing the ``mcount'' R package). 

% from section 2 to section 3, it is abrupt. Need to say why a simulation study is necessary at the end of section 2. what are we trying to achieve? That is the most critical piece. 1) I think we need to explain that sometimes interventions are targeted for populations that differ in their risk .... etc. 2) also need to say that there is no study that provides evidence-based guidance systematically for designing and planning future intervention trials (sample size) and also for analyzing data (Poisson - almost always wrong). ZZ: I feel this was explained in introduction?

\section{Simulation}
We conducted a simulation study to evaluate comparative performances across the following five statistical models in terms of empirical statistical power and Type I error in various data situations. 
\begin{enumerate}
\item [1.] Poisson model - testing the effect of intervention on the overall mean of the entire population ($H_0: \beta_1^{Poi}=0$).
\item [2.] NB model - testing the effect of intervention on the overall mean of the entire population ($H_0: \beta_1^{NB}=0$).
\item [3a.] ZIP model - testing the effect of intervention on the mean of the Poisson part ($H_0: \beta_1^{ZIP}=0$).
\item [3b.] ZIP model - testing the effect of intervention on the structural zero part ($H_0: \gamma_1^{ZIP}=0$).
\item [4.] MZIP model - testing the effect of intervention on the overall mean of the entire population ($H_0: \beta_1^{MZIP}=0$).
\item [5a.] Linear model with raw scale scores - testing the effect of intervention on the overall mean of the entire population ($H_0: \beta_1^{linear\_raw}$, where $\beta_1^{linear\_raw}$ is the intervention effect in the model).
\item [5b.] Linear model with log-transformed outcome scores - testing the effect of intervention on the overall mean of the entire population ($H_0: \beta_1^{linear\_log}$, where $\beta_1^{linear\_log}$ is the intervention effect in the model). \textcolor{black}{Note that a constant of 1 was added to outcome values to avoid taking logarithm on zero.} 
\end{enumerate}
Note that the ZIP model evaluates the intervention effect in two distinct parts, which come with two separate statistical tests for the intervention effect (i.e., Models 3a \& 3b). 

The simulation settings for study characteristics were based on motivating data from Project INTEGRATE \citep{mun2015project}, a large-scale individual participant data meta-analysis project examining the effectiveness of brief alcohol interventions on reducing alcohol consumption among young adults. Therefore, the simulation settings used in this study represent a broad range of data conditions in this field. Because the ZIP regression model allows for the flexible manipulation of the intervention effect on the two subpopulations through the Poisson and structural zero parts, it was selected as the data generating model. Specifically, for an individual study with two arms, we considered total sample sizes of $N\in\{100, 200, 300, 500\}$, where the outcome of the $i$-th subject ($i\in\{1,2,...,N\}$) was simulated by a ZIP model characterized by $y_{i}\sim \text{Poisson($\mu_{i}$)}$ with probability $1-\pi_{i}$, and 0 otherwise. The Poisson mean parameter $\mu_{i}$ and the structural zero rate $\pi_{i}$ were determined through the following link functions
\begin{equation}
\label{eq:simu}
\begin{split}
\log(\frac{\pi_i}{1-\pi_i})=\gamma_{0}+\gamma_{1}\mathbbm{1}_{\{A_{i}=T\}}+\gamma_{2}\text{Cov}_{i} \text{, and} \\
\log(\mu_{i})=\beta_{0}+\beta_{1}\mathbbm{1}_{\{A_{i}=T\}}+\beta_{2}\text{Cov}_{i},
\end{split}
\end{equation}
where the intervention group assignment was determined by $\mathbbm{1}_{\{A_{i}=T\}}\sim\text{Bernoulli}(0.5)$ and the covariate was generated by $\text{Cov}_{i}\sim N(0,1)$. Equation (\ref{eq:simu}) is a special case of the general formulation of a ZIP model (Equation (3)). For this simulation, we considered the situation, in which the outcomes are explained by the difference in the intervention vs. control group and an additional individual-level covariate for baseline differences in the outcome. 

The regression coefficients $\beta_1$ and $\gamma_1$ in Equation (\ref{eq:simu}) measure the intervention effects on the Poisson and structural zero parts of the ZIP model, respectively. In the context of alcohol intervention studies, $\beta_1$ could quantify the effect of the intervention on the average number of drinks for participants who drink, with $\beta_1<0$ representing a favorable intervention effect (i.e., reduced drinking). $\gamma_1$ would quantify the intervention effect on the proportion of abstainers, with $\gamma_1>0$ indicating a favorable intervention effect (i.e., a greater proportion of non-drinking) in the intervention arm. Since the intervention influences the outcome in two ways, we considered the following four intervention conditions characterized by different values of $\beta_1$ and $\gamma_1$:
% can we simply make a table to show all simulation conditions at a glance? 7 different model parameters (since zero portion part is the same in two models) x 4 intervention effect conditions x 7 zero rates = 196 x 1000 = 196,000 
\begin{itemize}[leftmargin=1in,label=\color{blue}\theenumi]
\item [Condition 1.] $\beta_1 \in \{-0.1,-0.2,-0.3\}$ and $\gamma_1 = 0.5$: Intervention \textit{reduces} the average number of drinks for those who drink (i.e., RR = 0.90, 0.82, 0.74) and \textit{increases} the proportion of abstainers or nondrinkers (i.e., OR = 1.65).
\item [Condition 2.] $\beta_1 \in \{-0.1,-0.2,-0.3\}$ and $\gamma_1 = 0$: Intervention \textit{reduces} the average number of drinks for those who drink (i.e., RR = 0.90, 0.82, 0.74) but \textit{has no effect on} abstainers (i.e., OR = 1).
\item [Condition 3.] $\beta_1 = 0$ and $\gamma_1 = 0.5$: Intervention \textit{has no effect on} the average number of drinks (i.e., RR = 1) for those who drink but \textit{increases} the proportion of abstainers (i.e., OR = 1.65).
\item [Condition 4.] $\beta_1 = 0$ and $\gamma_1 = 0$: Intervention \textit{has no effect on} both the average number of drinks and the proportion of abstainers (i.e., RR = 1 and OR = 1).
\end{itemize}
Conditions 1--3 will be used to evaluate statistical power across the models, where intervention is effective for the number of drinks, the likelihood of drinking, or both. In Condition 4, the intervention has no effect. This condition can be used to evaluate the Type I error rate of all models considered. 

To evaluate statistical properties over different levels of zero inflation, we varied the proportion of zero outcomes from 0.2, 0.3, ..., 0.8. \textcolor{black}{Once $\beta_1$ and $\gamma_1$ are determined in each condition, we set $\beta_0=0.8-\beta_1$ and $\beta_2=0.2$, so that the mean expected number of drinks is fixed, ensuring that samples are comparable across simulation conditions with respect to their drinking level. We further constrained  $\gamma_0=2\gamma_2$, so $\gamma_2$ can be calculated such that the pre-determined zero rate can be reached. Note that $\beta_2$ and $\gamma_2$ are the parameters quantifying the effects of a covariate in the simulation, which are of less interest in applied research.} The simulation settings are summarized in Table 1. In each simulation setting, we considered 1,000 replications. In each replication, study data were generated from a ZIP model based on the specified data condition explained above, and the models described in Section 2 were fit to evaluate rejection rates of the intervention effect. After 1,000 replications per condition, we calculated the rejection rate, which is the proportion of replications that yielded statistically significant results for the intervention for each condition for each model. In Conditions 1--3, which have at least one true intervention effect, the observed rejection rate is the empirical statistical power (i.e., true positives). In the final Condition 4 with null intervention effects, the rejection rate is the empirical Type I error rate, which is the ability to control the probability of having false positive results. We set the significance level at 0.05, so the rejection rate of 0.05 means that the method adequately controlled the Type I error. If the rejection rate is less than 0.05, the method is overly conservative in controlling the Type I error, which could lead to the inability to detect a true intervention effect, when it exists. If the rejection rate is greater than 0.05, the method may be prone to false positive results. The relative performance across the methods was evaluated by comparing their rejection rates from the models in each of the simulation settings.% \textcolor{green}{It maybe out of scope here. But for ZIP model, an overall treatment effect can be tested by log likelihood ratio test (with df=2) simultaneously with/without the treatment indicator in the structural zero and poisson parts. The downside is that LR test does not lead to point estimates on treatment effect. But in terms of type I error/power for the LR, I suspect it will be similar to the MZIP model. Maybe we can add into the discussion section.}

\section{Results}

\subsection{Type I error}
Figure 1 presents the rejection rates of the five statistical methods under null effects (Condition 4) across simulation settings. Since the intervention has no effect on the count and zero parts  ($\beta_1=\gamma_1=0$), the rejection rates represent the empirical Type I error. From the results, we see that first, the Poisson model failed to control the Type I error rate, which was highly inflated ($>0.05$) across simulation settings. Notably, as the zero rate increased, the inflation of the Type I error rate became more severe, indicating that the Poisson model is increasingly likely
to lead to false conclusions. Second, the MZIP model, the ZIP model testing the count part, and both of the linear models controlled the Type I error rate well ($\approx 0.05$) across simulation settings. Third, the MZIP and ZIP models testing the intervention effect on the structural zero part controlled Type I error well when sufficient zero inflation existed ($\ge 0.4$); however, when the zero inflation was modest (i.e., $\le 0.4$) and when the sample size was small to modest ($\le 300$), the tests became overly conservative with the Type I error rate less than 0.05. Fourth,  the NB model appropriately controlled Type I error ($\approx 0.05$) when the zero rate was low to moderate ($\le 0.3$). However, when the zero rate increased to 0.4 or higher, the Type I error rate fell below 0.05. Although the NB model is still valid in terms of statistical significance and its ability to control the probability of false positives, it is more likely to result in excessive false negative results under zero inflation. 
\subsection{Statistical power}
The rejection rates for the statistical methods under Conditions 1--3 are presented in Figures 2--4, respectively. Since the intervention had effects on at least the count or zero parts ($\beta_1 \text{ or } \gamma_1 \ne 0$), the rejection rates represent
empirical statistical power. The comparative results between the statistical methods for each of Conditions 1--3 are described. Note that the Poisson model will not be discussed here because it is statistically invalid under zero inflation (Figure 1) and exhibited the highest rejection rates (Figures 2--4) across the simulation settings.

\subsubsection{Condition 1: $\beta_1\in\{-0.1, -0.2, -0.3\}$ and $\gamma_1 = 0.5$}
In Condition 1, the intervention is efficacious for both the count and zero parts. First, as shown in Figure 2, the MZIP model testing the intervention effect on the overall mean and the linear model with raw scale scores had comparable statistical power, which was generally the highest. Second, the linear model on log transformed scores was less powerful than the MZIP and the linear model with raw scores under higher effect size ($\beta_1 = -0.2$ and $-0.3$), but the power disadvantage diminished at greater zero rates. When the effect size was small ($\beta_1 = -0.1$), the linear model on log transformed scores had the highest power at high zero rates ($>0.5$), but the power gain was modest compared to the MZIP model testing the overall mean and the linear model on the raw scale. Third, the NB model performed well at lower zero rates ($\le0.3$). However, as the zero rates increased, the statistical power of the NB model deteriorated quickly, showing much lower rejection rates than MZIP or linear models with either raw or log-transformed scores. This observation was expected because of the overly conservative Type I error rate of the NB model under moderate to high levels of zero inflation, which compromised its power to detect true intervention effects. Fourth, the ZIP model testing the count part had less power since testing the intervention effect only for the count part of the population leaves the intervention effects on the zero mass left ignored, leading to power loss. 

Notably, under a sample size of $N=100$, none of the methods reached a power of 0.8 in all simulation conditions. When $N=200$, a power of 0.8 was only achieved by the MZIP model, linear models, and NB models when the intervention effect on the mean was $-0.3$ (i.e., RR = 0.74) with low to moderate zero inflation. With a sample size of $N = 300$ or greater, statistical power was adequate for the MZIP and linear models in more simulation conditions with a zero rate as a major determining factor of power, along with the magnitude of effects. 
% this reminds me that we want to do interactive display like Huh et al. 
\subsubsection{Condition 2: $\beta_1\in\{-0.1, -0.2, -0.3\}$ and $\gamma_1 = 0$}
Suppose that intervention was efficacious only for participants who engaged in the behavior of interest, such as consuming drinks containing alcohol, but not for those who predictably did not drink. First, as shown in Figure 3, the ZIP model testing the Poisson mean outperformed the four other statistical models (i.e., MZIP, NB, and linear models with raw or log transformed scores). It may be because the other models could not leverage the intervention effect on the zero mass when estimating the overall treatment effect on the overall mean when $\gamma_1 = 0$, and also because it was the ``true" model. Second, the MZIP model testing the intervention effect on the overall mean had the highest power of the remaining four tests, followed by the linear model with raw scores, the NB model, and the linear model with log-transformed scores. %Similar to the simulation results in Figure 2, statistical power was directly influenced by sample size, effect size, and zero rate.  

% when gamma_1 is zero, why the bottom two lines are elevated from zero? especially for low sample size conditions and more so for higher zero rates? Perhaps we can mention that because there is very little data at high zero rates.
\subsubsection{Condition 3: $\beta_1=0$ and $\gamma_1 = 0.5$}
Finally, in Condition 3, intervention had an effect only on the proportion of zero responses (e.g., abstainers). Figure 4 shows the results on the power to detect intervention effects. First, as the ``true" model in this condition,
the ZIP model testing the  structural zero part had the highest power with larger samples ($N\ge300$). The linear model  with log-transformed outcome scores generally had the second highest power, followed by the linear model with raw scores and the MZIP model. Moreover, even at $N=500$, the power to detect the intervention effect was less than 0.8. For $N=300$ or less, statistical power was below 0.5. Of note, we anticipate that this condition is less common in practice as interventions generally tend to influence the average number of drinks if they affect the likelihood of abstaining. Therefore, the result of Condition 3 is not our primary interest in the current study. %Note that under sample sample sizes, they had lower power than the linear model on log scale, which may be due to a loss of estimation efficiency with insufficient sample size. Second, the linear model on log scale had slight power loss compared to tests focusing on structural zeros, followed by the MZIP model on overall mean, linear model on raw scale, NB model, and the ZIP model testing Poisson mean, which had almost zero power.
\section{Discussion and Conclusions}

We conducted an extensive simulation study to evaluate and compare the statistical properties among candidate methods for count data in the context of health behaviors research. We represented a variety of plausible data situations in terms of the size and direction of the effect, sample size, and degree of zero inflation. The empirical results obtained from this simulation can serve as a proxy to real data applications in the field. We provide clinical implications and practical recommendations for model selection.

Among the conventional count distribution-based models without adjustment for zero-inflation, the Poisson model is always invalid with an inflated Type I error rate under zero inflation (e.g., zero rate $\ge$ 0.2). The Poisson model tends to falsely judge an ineffective intervention to be efficacious, leading to excessive false positive results. This result can be expected because the Poisson model does not allow modeling excessive zeros nor overdispersion, which typically occurs in zero-inflated data \citep{yang2009testing}. Therefore, we recommend against using the Poisson model in all scenarios where zero inflation is present. Similar to the Poisson model, the NB model does not account for excessive zeroes, but allows for overdispersion. However, it controls the Type I error below the nominal level (e.g., $\alpha = 0.05$). When zero inflation is moderate to high (e.g., zero rates $\ge$ 0.4), the NB model tends to overly control Type I error, hampering one's ability to detect true intervention effects. Although the NB model is still statistically valid under zero inflation, its statistical power is compromised.

The ZIP model is an extension of the Poisson model that accounts for excessive zeros through a mixture distribution of the Poisson and a point mass at zero %It has the highest power when any part of the ZIP distribution, that is, the Poisson mean or structural zero rate, is of interest, which are explicitly modeled via the two-part, ZIP model. In health behavior research, the Poisson mean and structural zero rate correspond to two distinct subpopulations, which are those who may or may not engage in the behavior of interest (e.g., drinking) at a given assessment, and those who ``predictably'' do not engage, respectively. Therefore, the choice of the ZIP model is relevant to the trial's population of interest. In an alcohol intervention trial for college students who regularly drink, investigators may be more interested in students who may potentially drink in social events and less interested in those abstain from drinking, thus the ZIP model testing the Poisson mean may be preferred because of its focus on that particular subpopulation. %In another study aimed at controlling initiation of alcohol use among first-year college students who do not drink before,    
and is most powerful when the main interest is any single part of the ZIP distribution. However, when the overall intervention effect on the entire population is of interest, it has less power than the MZIP model. When the intervention has favorable effects on both the subpopulations (i.e., reducing the average number of drinks among those who may drink and increasing the proportion of abstainers), the MZIP model generally had superior statistical power. This is because the MZIP model evaluates the effects on the overall mean of the outcome directly, which combines the effects from both subpopulations, yielding higher statistical power. Of note, although the ZIP model was used to simulate the data, it may have less power  \textcolor{black}{because the tested parameters, $\beta_1^{ZIP}$ and $\gamma_1^{ZIP}$, captured the information from one of the corresponding subpopulations. In contrast, the MZIP and other models were able to capture the information from the entire population.}

Under the studied simulation conditions, linear models are also valid with well-controlled Type I error rates. The linear model with raw scale scores generally had higher statistical power compared with its counterpart with log transformation. This observation suggests that although the use of count distribution-based models has been widely promoted for count data, the linear models may still produce valid inferences with acceptable statistical power, especially when compared with the Poisson or NB models. However, the use of log transformation may not be optimal for count outcome analysis with zero inflation due to information loss. For example, the count outcome of number of drinks consumed in a week has a relatively limited range (e.g., 0--30 drinks), so taking the log transformation may not be as beneficial as in other situations with a large range, such as expenses in dollars in economic research. Of note, the linear model with raw scale scores had almost identical statistical power to that of the MZIP model, which may be because both methods target the mean difference in the outcomes.

%If people want to use count based methods, better use the appropriate distribution, otherwise the perforamnce maybe bad - worse then linear models.

%Heterogenity of populations and within populations who are receiving treatment, mixture models have been alternative used (e.g., ...). But they are based on assumotions that the subpopulations are normally distribution, which may not be true. (leave it for now; for grant purpose)

%assumes all subpopulation to be the same... for ZIP and MZIP, they allow subpopukations to differ, which may be better. -- think about it for future directions 

\textcolor{black}{The simulation settings studied in the current study were motivated by the data from brief alcohol intervention studies. Therefore, the findings are most relevant for the alcohol intervention trials. 
However, many behavioral interventions with count outcomes have similar effect sizes. Thus, the findings reported may have a broad impact beyond the motivation data.} With zero-inflated outcome data, we also observed that statistical power across the methods was mostly below 0.8 under small to moderate sample sizes (e.g., $N\le300$) and effect sizes (e.g., $|\beta_1|\le$ 0.2, corresponding to an intervention effect that reduces no more than 19\% of the average number of drinks among participants who may drink). Even at a large sample size of $N = 500$, the power was adequate (at the 0.8 level) only for a few conditions.  Specifically, typical clinical trials would not have adequate power to detect a significant intervention effect under small to moderate effect sizes with small to moderate samples, regardless of the statistical methods used. Our findings indicate that in the presence of zero inflation, individual clinical trials lack statistical power to detect effects on count outcomes. This underscores the value of meta-analysis using individual participant data for increasing statistical power when analyzing such data (e.g., \cite{mun2022brief}). Study-specific small or null effects can be combined together to provide large-scale robust evidence in the field of brief alcohol intervention and related areas (e.g., \cite{mun2015project}).

\section*{Declarations}

\subsection*{Ethics approval and consent to participate}
Not applicable.

\subsection*{Consent for publication}
Not applicable.

\subsection*{Availability of data and materials}
The program code of the MZIP method is available from the R package ``mcount" \citep{mcount}. The script for simulation and analysis is available at Mendeley Data (https://data.mendeley.com/datasets/r5bztdd766; \citep{zhou2023data}).

\subsection*{Competing interests}
The authors declare that they have no competing interests.

\subsection*{Funding}
The project described was supported by the National Institute on Alcohol Abuse and Alcoholism (NIAAA) grants R01 AA019511 and K02 AA028630 and the National Science Foundation (NSF) grants DMS2015373 and DMS2027855. The content is solely the responsibility of the authors and does not necessarily represent the official views of the NIAAA, the National Institutes of Health, or the NSF.

\subsection*{Authors' contributions}
ZZ, DL, and EYM initially wrote the manuscript’s early drafts. ZZ and DL prepared the computing script, and ZZ conducted the simulation and prepared all figures and tables. DH helped conceptualize a methodological gap in clinical applications. EYM and MX provided feedback on the design and interpretation of the simulation study. All authors reviewed and edited earlier drafts and approved the final version of the manuscript.

\subsection*{Acknowledgements}
Not applicable.

\bibliography{reference}

\bibliographystyle{agsm}

\newpage
% \usepackage{multirow}
\begin{table}[H]
\centering
 \caption{Summary of simulation settings}
 \vspace{0.3cm}
\begin{tabular}{|l|l|}
\hline
            & Possible values     \\ \hline
Sample size & 100, 200, 300, 500  \\ \hline
Zero rate   & 0.2, 0.3, ..., 0.8  \\ \hline
$\beta_1$       & 0, -0.1, -0.2, -0.3 \\ \hline
$\gamma_1$    & 0, 0.5              \\ \hline
\end{tabular}
\end{table}

\newpage

\begin{figure}
\begin{center}
\includegraphics[width=7.5in]{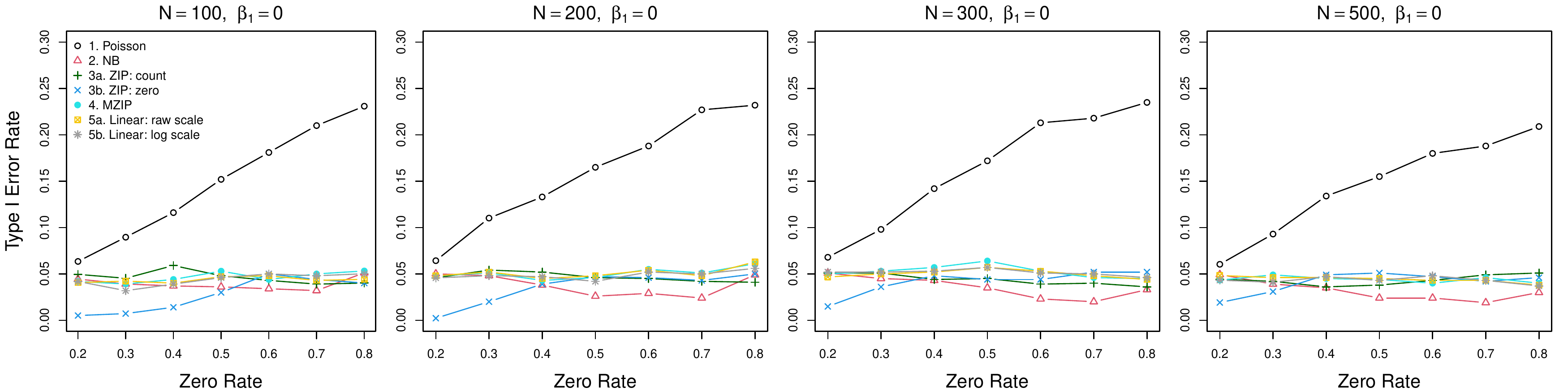}
\end{center}
\caption{Results of empirical Type I error rates under Condition 4 for different statistical methods from 1000 replications  \label{fig:1}}
\end{figure}

\begin{figure}
\begin{center}
\includegraphics[width=6.5in]{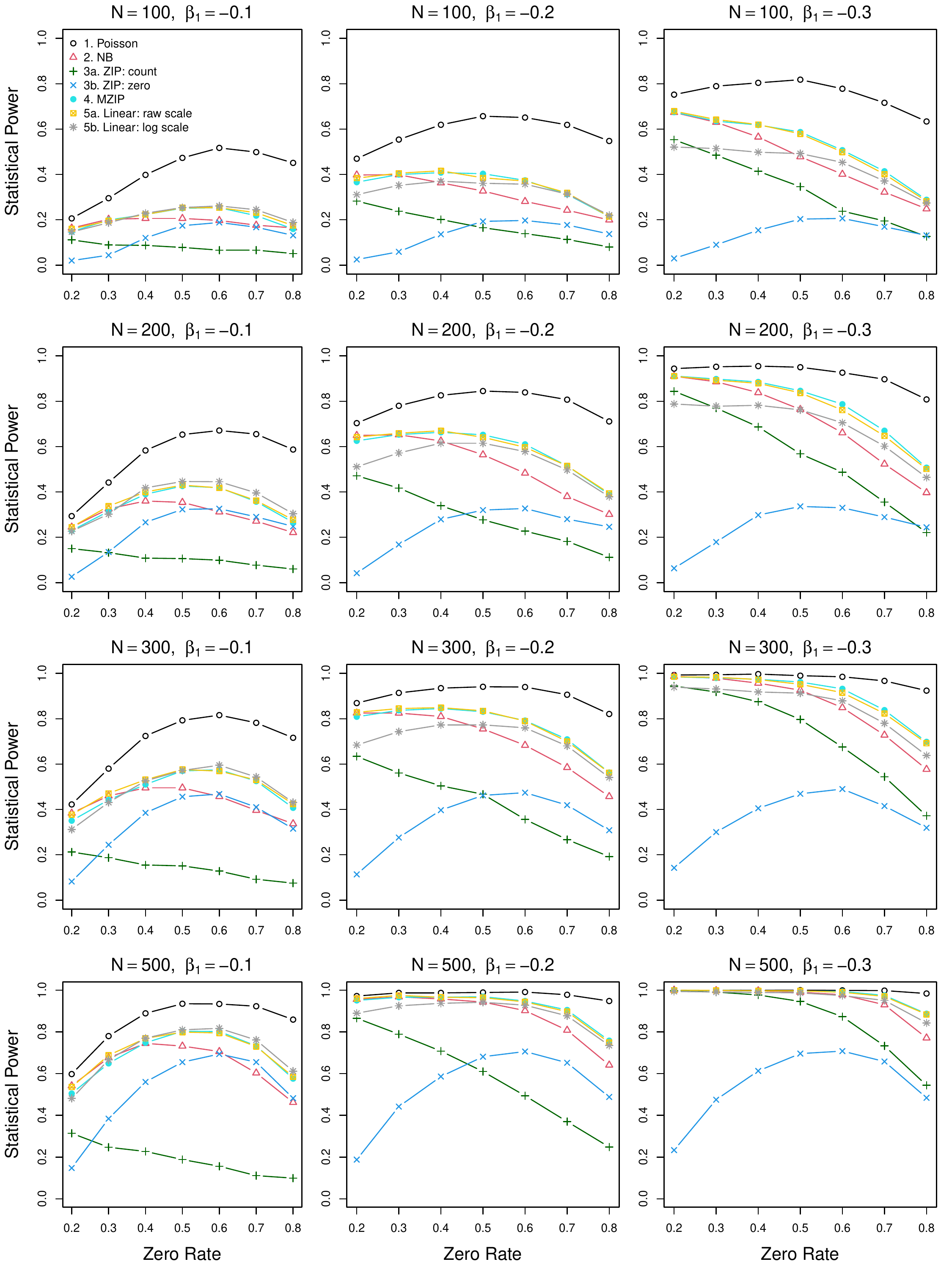}
\end{center}
\caption{Results of empirical statistical power under Condition 1 for different statistical methods from 1000 replications  \label{fig:2}}
\end{figure}

\begin{figure}
\begin{center}
\includegraphics[width=6.5in]{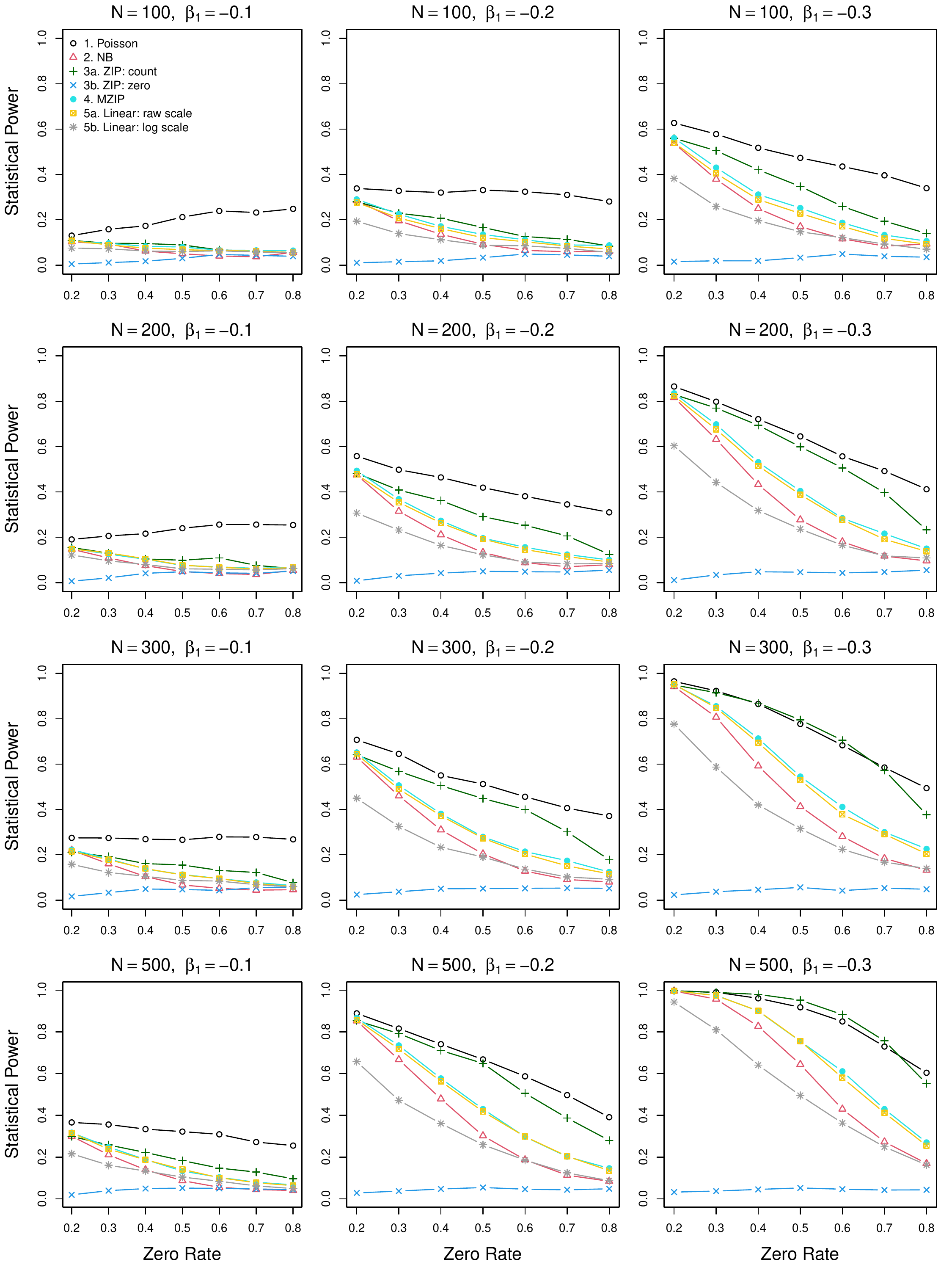}
\end{center}
\caption{Results of empirical statistical power under Condition 2 for different statistical methods from 1000 replications \label{fig:3}}
\end{figure}

\begin{figure}
\begin{center}
\includegraphics[width=7.5in]{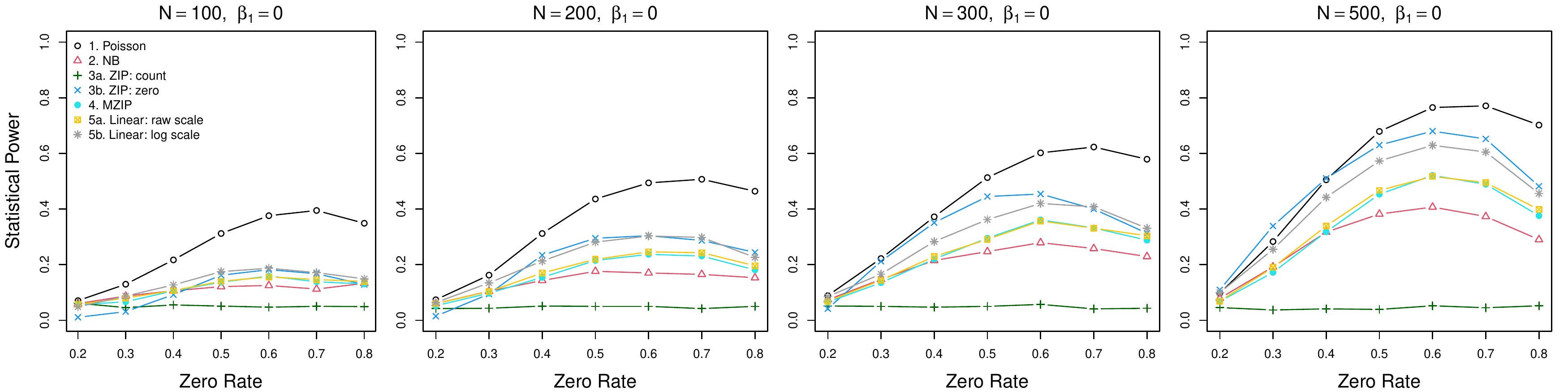}
\end{center}
\caption{Results of empirical statistical power under Condition 3 for different statistical methods from 1000 replications  \label{fig:4}}
\end{figure}

\newpage

\end{document}